\documentclass[intlimits,twoside,a4paper]{article}
\usepackage{amsmath,amssymb}
\usepackage{graphicx}
\usepackage{wrapfig}
\usepackage[T2A]{fontenc}
\usepackage[cp1251]{inputenc}
\usepackage{footmisc}
\usepackage{bm}

\usepackage{threeparttable}

\usepackage[eqsecnum]{cmpj}

\renewcommand{\Im}{\mathfrak{J}}

\issue{2011}{14}{3}{33002}

\doinumber{10.5488/CMP.14.33002}

\title[Molecular mechanism of surface charge]%
{On the molecular mechanism of surface charge amplification and related phenomena at aqueous polyelectrolyte-graphene interfaces%
\thanks{The submitted manuscript has been authored by the contractor of the U.S. Government under contract No.~DE--AC05--00OR22725. Accordingly, the U.S. Government retains a nonexclusive, royalty-free license to publish or reproduce the published form of this contribution, or allow others to do so, for U.S. Government purposes.}}

\author[A.A. Chialvo, J.M. Simonson]{A.A. Chialvo\refaddr{label1}\thanks{E-mail: chialvoaa@ornl.gov, FAX 865-574-4961}\,, J.M. Simonson\refaddr{label2}}
\addresses{
\addr{label1}Chemical Sciences Division Geochemistry and Interfacial Sciences Group,
Oak Ridge National Laboratory, Oak Ridge, TN 37831--6110, U.S.A.
\addr{label2}Neutron Scattering Sciences Division, Oak Ridge National Laboratory,
Oak Ridge, TN 37831--6475, U.S.A.}

\authorcopyright{A.A. Chialvo, J.M. Simonson, 2011}
\date{Received April 11, 2011, in final form May, 18, 2011}

\begin{document}

\maketitle

\begin{abstract}
In this communication we illustrate the occurrence of a recently reported new phenomenon of surface-charge ampli\-fi\-cation, SCA, (origi\-nally dubbed overcharging, OC), [Jimenez-Angeles~F. and Lozada-Cassou~M., J. Phys. Chem.~B, 2004, {\bf 108}, 7286] by means of molecular dynamics simulation of aqueous electrolytes solutions involving multivalent cations in contact with charged graphene walls and the presence of short-chain lithium polystyrene sulfonates where the solvent water is described explicitly with a realistic molecular model.  We show that the occurrence of SCA in these systems, in contrast to that observed in primitive models, involves neither contact co-adsorption of the negatively charged macroions nor divalent cations with a large size and charge asymmetry as required in the case of implicit solvents.  In fact the SCA phenomenon hinges around the preferential adsorption of water (over the hydrated ions) with an average dipolar orientation such that the charges of the water's hydrogen and oxygen sites induce magnification rather than screening of the positive-charged graphene surface, within a limited range of surface-charge density.
\pacs 07.05.Tp, 61.20.Ja, 82.35.Rs, 68.08.-p, 61.20.Ja
\end{abstract}

\section{Introduction}
\label{sec:Introduction}

The conventional view on the behavior of electrolyte solutions in contact with a charged surface hinges around the concept of surface-charge screening by the ions in solution, which ultimately renders the entire system electroneutral (for a detail discussion on the surface-charge screening phenomenon the reader should refer to references~\cite{Jimenez1,Sjostrom2,hansen3}).  This screening might take the form of adsorption, i.e., by alternating layers of opposite charged ions that gives rise to an oscillatory behavior for the total axial charge-density and represented by charge reversal (CR) followed by charge inversion (CI) phenomena. Note that for primitive models, such as the Poisson-Boltzmann (PB) description, this (apparent surface charge or local net charge density $\sigma(z)$, {vide infra}) profile is actually exponentially monotonous, and consequently, it never changes sign~\cite{Sjostrom2}.

CR and CI phenomena are obviously the result of the stacking of alternately charged species that manifest as non-monotonous (oscillatory) profiles for the local electric field and potential i.e., they depend on ion-ion correlations or they are correlation-induced phenomena~\cite{hansen3}) whose most obvious practical application is the layer-by-layer deposition of polyelectrolyte toward membrane formation~\cite{Decher4}.

Jimenez-Angeles and Lozada-Cassou~\cite{Jimenez1} have recently reported a new phenomenon, dubbed overcharging, that occurs during the adsorption of macroions in solution onto planar charged-surfaces, i.e., when the macroions bring with them counterions to the surface that contribute to the buildup of a like charge.  This phenomenon was initially revealed by integral equation calculations of three-component inhomogeneous primitive models of macroion solutions in contact with a positive-charged surface, and represented the adsorption of ``an effective charge onto a like-charged'' surface.  In fact, when referring to this new phenomenon the authors indicated that ``\emph{such an effect defines a new phenomenon, hereafter referred to as overcharging (OC), i.e., at the wall's neighborhood we find the accumulation of an effective additional charge with the same sign of the wall.  This effect is due to the strong electrostatic attraction between macroions and the divalent cations.  However, for this effect to be present, a high particle's excluded volume is needed, i.e., a high concentration of macroions and/or little particles and/or large macroion size or little ion size}''.

Other authors have reported OC phenomena since the original work of Jimenez-Angeles and Lozada-Cassou~\cite{Jimenez1}, involving either bulk~\cite{GarciaSoft5,GarciaJour6} or interfacial~\cite{Messina7,Wang8,Wang9} macroion solutions.  A common denominator in all these studies~\cite{Jimenez1,GarciaSoft5,GarciaJour6,Messina7,Wang8,Wang9,Yu10} is the involvement of primitive models that ignore the discrete molecular-based nature of the solvent, and therefore, they neglect among other things the excluded volume of the solvent and the solvation effects that translate into significant dielectric inhomogeneous environments around the charged species.

The goal of this communication is to illustrate, by molecular dynamics simulation, the occurrence of OC in model aqueous electrolytes solutions involving multivalent cations in contact with positive-charged graphene walls in the presence of short-chains of lithium polystyrene sulfonate, where all species including the solvent water are explicitly and atomistically described.  We argue that the occurrence of OC in these explicit solvent systems involves the preferential adsorption of the water's hydrogen and oxygen sites rather than the co-adsorption of the negative-charged macroions and the divalent cations with a large size and charge asymmetry, as observed in the primitive systems.

To address this issue, we perform extensive molecular dynamics simulations on similar systems to those described in our previous work~\cite{Chialvo11}, over a wider range of positive surface charges as described in section~\ref{sec:Fundamentals}. In section~\ref{sec:Discussion} we present and discuss the relevant axial profiles to identify the molecular mechanism underlying the OC mechanism in these systems.  Finally, in section~\ref{sec:Concluding} we discuss the contrasting differences between the OC phenomenon involving implicit and explicit description of the solvent water.

\section{Fundamentals and simulation approach}
\label{sec:Fundamentals}

The word ``overcharging'' has been most often used to describe the process of ``overcompensation'' of the surface charge on a solid substrate or a macroion due to the adsorption of multivalent counterions (e.g., references~\cite{Gossl12,Grosberg13}), a phenomenon that has been known for a rather long time under alternative names including ``charge inversion''~\cite{Grosberg13,Joanny14}, ``charge overcompensation''~\cite{Dobrynin15}, and ``charge reversal''~\cite{Diehl16}.  However, here we are dealing with a relatively new phenomenon, first reported by Jimenez-Angeles and Lozada-Cassou~\cite{Jimenez1}, involving the ``amplification'' of the surface charge, i.e., the actual increase of the original surface charge by the co-adsorption of macro- and micro-ions.  In what follows we will invoke Jimenez-Angeles et al.'s definition of OC that describes the adsorption of an effective charge onto a like-charged wall (see also italicized quote vide supra), though we will rather refer to as the surface charge amplification (SCA) phenomenon~\cite{GarciaSoft5,Messina7} to avoid any confusion with related phenomena, and illustrate how the differential hydration behavior of hydrated ions and polyions in conjunction with the significant graphene-water interactions contributes to its occurrence.

To study the SCA in the current systems we place emphasis on the determination of the axial distribution profiles that characterize the structure of the graphene-aqueous interface, including all species concentrations $\rho_i(z)$, the corresponding Coulombic charge density $\rho_Q(z)$, and the relative orientation of the water molecules $\theta(z)$ as follows~\cite{Chialvo11},
\begin{equation} \label{eq1}
{\rm P} (z)\equiv \left\langle \left(L_{x} L_{y} \Delta \right)^{-1} \sum _{i}{\rm P} _{i} {\kern 1pt} {\rm B} _{\rm c} \left(z_{i}\, ,-0.5\Delta ,+0.5\Delta \right) \right\rangle
\end{equation}
for which $P_i$ denotes $\delta(z - z_i)$, the Coulombic charge $q_i$\,, and the angle $\theta =\cos ^{-1} \left({{\bm \mu }_{{\it \iota }}
\cdot\hat{z}\mathord{\left/ {\vphantom {{\bm \mu }_{{\it \iota }} \bullet \hat{z} \left|{\bm \mu }_{{\it \iota }} \right|}} \right. \kern-\nulldelimiterspace} \left|{\bm \mu }_{{\it \iota }} \right|} \right)$ between the water's dipole moment ${\bm \mu}_i$ and the unit vector $\hat{z}$ perpendicular to the graphene surface, respectively, while ${\rm B} _{\rm c} \left(x,a,b\right)\equiv \left[\Theta \left(x-a\right)-\Theta \left(x-b\right)\right]$ defines the ``boxcar'' function~\cite{vonSeggern17}, $L_{\alpha } $ represents the size of the simulation box along the $\alpha$-axis, $\Delta \sim 0.3$~\AA, and the $\left\langle \cdots\right\rangle$ indicates a time average over the simulation trajectory.  Note that the first two distributions are connected by the expression $\rho _{Q} (z)=\sum _{i}x_{i} q_{i} \rho _{i} (z) $, the one that we use for a test of internal consistency.

Subsequently, the axial integration of $\rho _{Q} (z)$ according to the Poisson equation provides us with a way to assess the strength of the (normalized) surface-charge screening, $\Im (z)\equiv 1-\left({\sigma (z)\mathord{\left/ {\vphantom {\sigma (z) \sigma _{\rm s} }} \right. \kern-\nulldelimiterspace} \sigma _{\rm s} } \right)$, where the local net (surface) charge density $\sigma (z)$ is given by,
\begin{equation} \label{eq2}
\sigma (z)=\sigma _{\rm s} +\int _{0}^{z}\rho _{Q} (z)\, \rd z
\end{equation}
subjected to the electroneutrality condition $\sigma (h)=0$, i.e.,

\begin{equation} \label{eq3}
\sigma _{\rm s} =-\int _{0}^{h}\rho _{Q} (z)\, \rd z.
\end{equation}

Consequently, according to Gauss' law, the profile of the electric field $E(z)$ corresponding to $\sigma (z)$ becomes,

\begin{equation} \label{eq4}
E(z)=4\pi \sigma (z)
\end{equation}
with $E(0)=4\pi \sigma (0)$, i.e., $\sigma (0)=\sigma _{\rm s}$\,, and from equations \eqref{eq2}--\eqref{eq4} it follows that,
\begin{equation} \label{eq5}
E(z)=-4\pi \int _{z}^{h}\rho _{Q} (z) \rd z.
\end{equation}

Now, we can reinterpret the strength of the surface-charge screening $\Im (z)$ as follows,
\begin{equation} \label{eq6}
\Im (z)=1-\left[{E(z)\mathord{\left/ {\vphantom {E(z) E(0)}} \right. \kern-\nulldelimiterspace} E(0)} \right]
\end{equation}
where equation \eqref{eq6} describes how effectively the aqueous electrolyte-polyelectrolyte screens the surface charge of the graphene wall.  While $\Im (z)$ satisfies two obvious conditions, i.e., null screening, $\Im (0)=0$, and full screening, $\Im (h)=1$, this function might not be necessarily bounded by those conditions due to the non-monotonous behavior of $\rho _{Q} (z)$. It is quite possible to find ({vide infra}) that the most interesting interfacial phenomena involving adsorption of macroions in the presence of aqueous electrolytes are associated with the condition $0>\Im (z)>1$.

In principle, the integrated fluid charge $\sigma (z)-\sigma _{\rm s} =\int _{0}^{z}\rho _{Q} (z) \rd z {\rm \; }$can exhibit an oscillatory behavior around electroneutrality, yet, for a positive-charged surface ($\sigma _{\rm s} >0$) the graphene-aqueous electrolyte interface would display three distinct types of behavior, depending on the strength of the local net (surface) charge density at its first extreme (peak or valley), {i.e}., $\sigma (z_{\otimes } )=\sigma _{\otimes } $\,, relative to the surface charge $\sigma _{\rm s}$\,, namely:

(a) surface-charge amplification (SCA), i.e.,
\begin{equation} \label{eq7}
\sigma _{\otimes } >\sigma _{\rm s} >0{\rm \; \; \; }\to {\rm \; \; }\int _{0}^{z_{\otimes } }\rho _{Q} (z)\rd z >0{\rm \; \; }\to {\rm \; }\Im (z_{\otimes } )<0;
\end{equation}

 (b) charge inversion (CI), i.e.,
\begin{equation} \label{eq8}
\sigma _{\rm s} >\sigma _{\otimes } >0{\rm \; \; \; }\to {\rm \; \; }\sigma _{\rm s} >-\int _{0}^{z_{\otimes } }\rho _{Q} (z)\rd z {\rm \; \; }\to {\rm \; }\Im (z_{\otimes } )<1;
\end{equation}

(c) charge reversal (CR), i.e.,
\begin{equation} \label{eq9}
\sigma _{\otimes } <0{\rm \; \; \; }\to {\rm \; \; }\sigma _{\rm s} <-\int _{0}^{z_{\otimes } }\rho _{Q} (z)\rd z {\rm \; \; }\to {\rm \; }\Im (z_{\otimes } )>1.
\end{equation}

Our systems comprised aqueous electrolyte solutions involving short-chain lithium polystyrene-sulfonate in contact with a positive-charged graphene surface, $0\leqslant \sigma _{\rm s} (\rm C \ m^{-2} )\leqslant 0.1525$, with added barium chloride to render the systems electroneutral.  Full details of the simulation methodology, description of the models, system sizes and relevant simulation parameters are given in our previous work~\cite{Chialvo11}, consequently, here we only provide the most relevant information regarding the system composition and surface charge in table~\ref{tab:surface}.

\begin{table}[h]
\begin{center}
\begin{threeparttable}
\caption{Surface charge and composition of aqueous polyelectrolyte \, solutions.}\label{tab:surface}
\begin{tabular}{|c|c|c|c|c|c|c|c|}\hline
System & $\sigma _{\rm s} /$C \ m$^{-2}$ & ${q_{\rm s}}^{(a)}$ & $N_{\rm Li^{+} } $ & $N_{\rm Ba^{2+} } $ & $N_{\rm Cl^{-} } $ & $N_{\rm H_{2} O} $ & $N_{\rm SO_{3}^{-} } $ \\ \hline
\hline
1 & 0.1525 & 30 & 100 & 20 & 90 & 4000 & 80 \\ \hline
2 & 0.1017 & 20 & 100 & 20 & 80 & 4000 & 80 \\ \hline
3 & 0.0763 & 15 & 100 & 20 & 75 & 4000 & 80 \\ \hline
4 & 0.0508 & 10 & 100 & 20 & 70 & 4000 & 80 \\ \hline
6 & 0.0254 & 5 & 100 & 20 & 65 & 4000 & 80 \\ \hline
7 & 0 & 0 & 100 & 20 & 60 & 4000 & 80 \\ \hline
\phantom{$^{ (b)}$}8$^{ (b)}$ & 0 & 0 & 0 & 0 & 0 & 4000 & 0 \\ \hline
\multicolumn{8}{l}{${}^{(a)}$\small Total electrostatic charge at carbon sites exposed to the fluid phase.}\\
\multicolumn{8}{l}{${}^{(b)}$\small Pure water in contact with the graphene wall. }\\
\end{tabular}
\end{threeparttable}
\end{center}
\end{table}

\section{Discussion of simulation results}
\label{sec:Discussion}



\begin{figure}[!h]
\vspace{-0.5cm}
\includegraphics[width=0.5\textwidth]{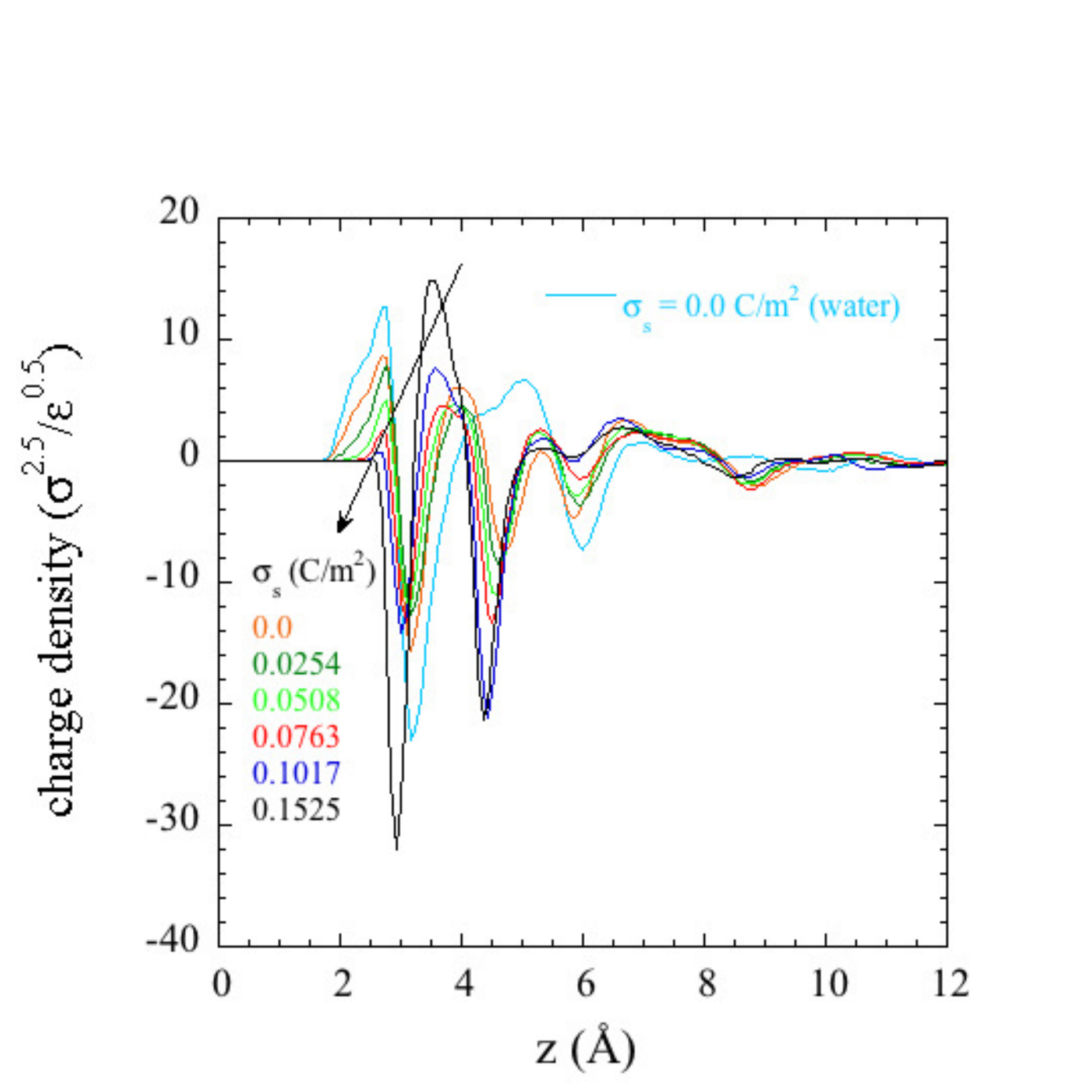}%
\hfill%
\includegraphics[width=0.5\textwidth]{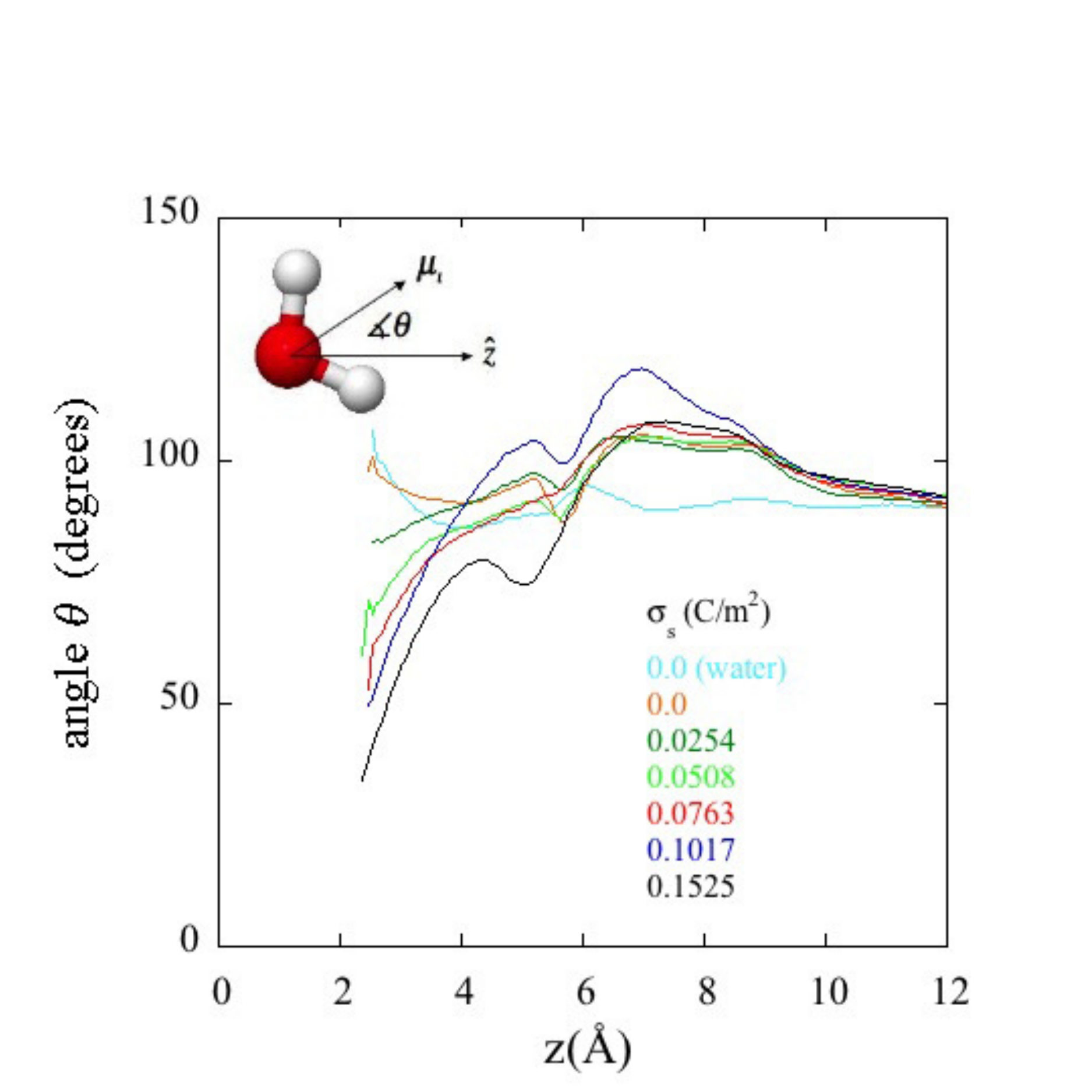}%
\vspace{-0.4cm}
\\
\parbox[t]{0.5\textwidth}{%
\caption{(Color on-line) Behavior of the axial profiles for the charge density, $\rho _{Q}
(z)$, for the graphene/fluid interface when the fluid is either pure water at
zero surface charge or an electrolyte-polyelectrolyte solution with a
surface-charge range of $0\leqslant \sigma _{\rm s} (\rm C~m^{-2} )\leqslant 0.1525$.  Charge
densities in units of SPC/E Lennard-Jones parameters $\sigma _{\rm OO} =3.166$~\AA\,
and $\varepsilon _{\rm OO}/k =78.23$~K.\label{charge}}
}%
\hfill%
\parbox[t]{0.5\textwidth}{%
\caption{(Color on-line) Behavior of the axial profiles for the relative orientation $\theta (z)$ of the water molecules in the graphene/fluid interface when the fluid is
either pure water at zero surface charge or an electrolyte-polyelectrolyte
solution with a surface-charge range of $0<\sigma _{\rm s} (\rm C~m^{-2} )\leqslant 0.1525$.\label{axial}}
}%
\end{figure}

In figure~\ref{charge} we display the axial profiles for the average charge
densities, corresponding to the surface-charge range of
$0\leqslant \sigma _{\rm s} (\rm C~m^{-2} )\leqslant 0.1525$, within the first 12~\AA\, from the graphene surface, in comparison with that corresponding to pure water in
contact with the uncharged graphene surface.  The first relevant feature in this
comparison is the appearance of an additional peak about 3.5--4~\AA \, from the
graphene surface, sandwiched between the original two in the pure-water
profile, originated in the adsorption of hydrated charged species (including
paired and condensed cations).  In addition, there is a strong reduction of
the size of the first peak of the charge density (located at $\sim$ 2.6~\AA)
with a simultaneous strengthening of the second peak (at $\sim$ 3.5--4~\AA) as
the surface-charge increases to the extent that the first peak disappears as
the surface charge approaches $\sigma _{\rm s} \cong 0.1525$~C~m$^{-2}$.  The
observed decrease of the strength of the first peak is accompanied by an
increase of the strength of the second peak and the deepening of the two
corresponding valleys (at $\sim$~3~\AA\, and $\sim$~4.5~\AA), a behavior that points
to the re-structuring of the interfacial region through the adjustment of
species adsorption and water re-orientation as we illustrate below.  In fact,
figure~\ref{axial} captures the evolution of the relative dipolar orientation
of the water molecules in the
interfacial region, where it becomes clear that the hydrogen sites of the water molecules in the
adsorbed layer of either a graphene/water or graphene/aqueous interface with
$\sigma _{\rm s} =0$ are tilted toward the graphene surface, i.e.,
$\theta >90^{\circ } $.  As we increase the surface charge, i.e.,
$\sigma _{\rm s} \to 0.1525$~C~m$^{-2}$, the positive-charged hydrogen sites are
repelled by the graphene charged surface, and consequently, $\theta \ll 90^{\circ } $.


\begin{figure}[!b]
\vspace{-0.8cm}
\includegraphics[width=0.5\textwidth]{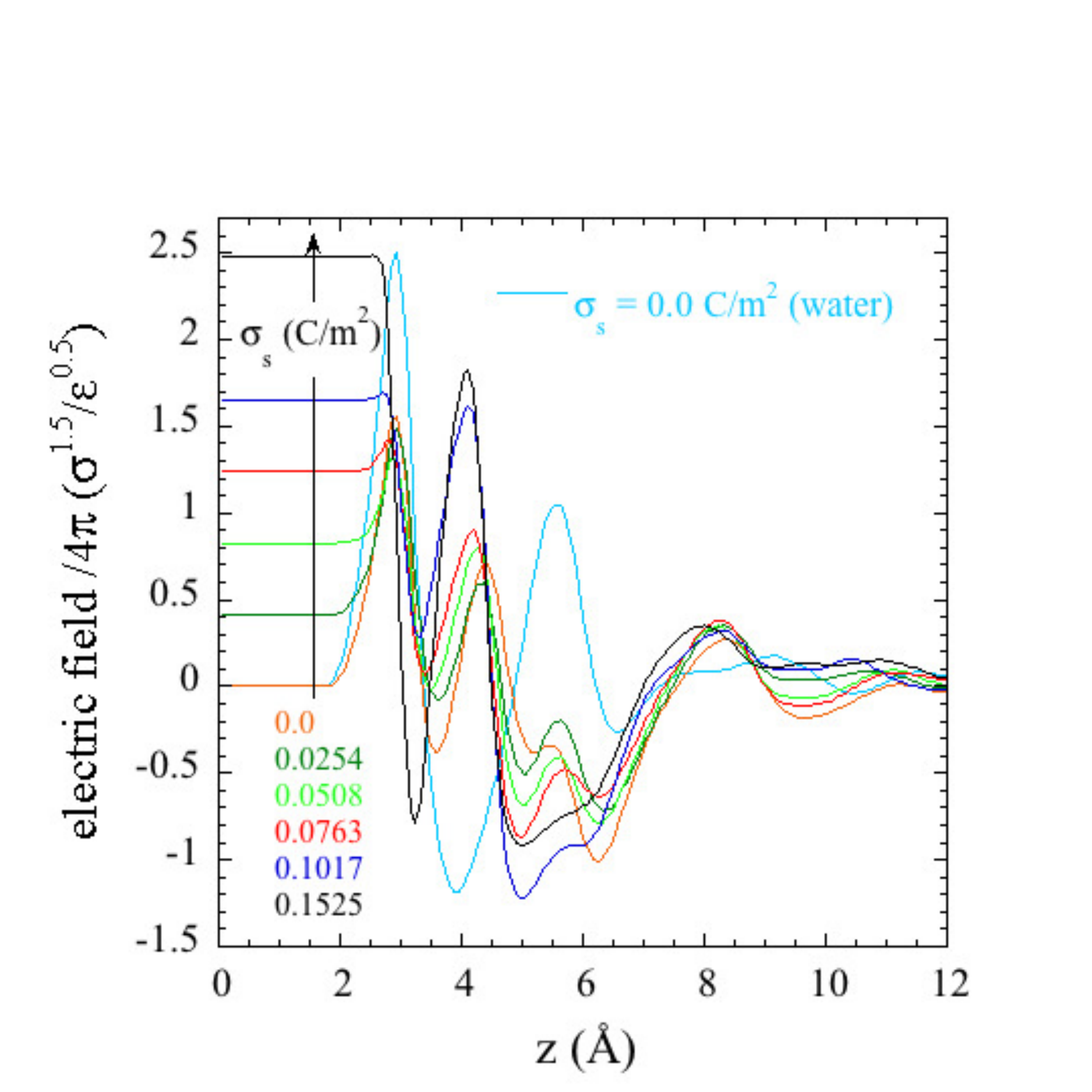}%
\hfill%
\includegraphics[width=0.5\textwidth]{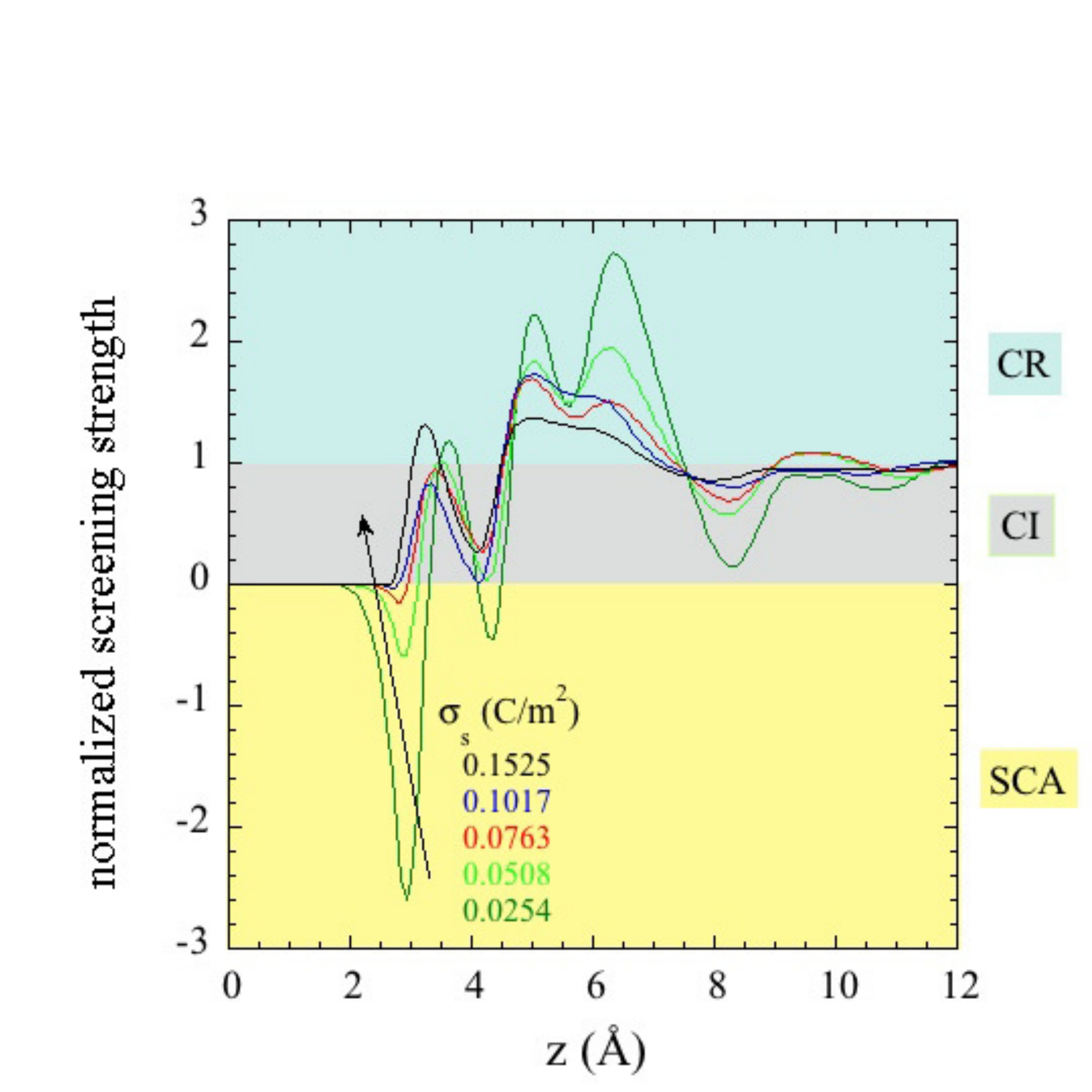}%
\vspace{-0.4cm}
\\
\parbox[t]{0.5\textwidth}{%
\caption{(Color on-line) Behavior of the axial profiles for the integrated fluid charge
density, $\sigma (z)=E(z) / 4\pi $, for the graphene/fluid interface when the
fluid is either pure water at zero surface charge or an
electrolyte-polyelectrolyte solution with a surface-charge range of $0\leqslant \sigma _{\rm s} (\rm C~m^{-2} )\leqslant 0.1525$.  Integrated charge densities in units of
SPC/E Lennard-Jones parameters $\sigma _{\rm OO} =3.166$~\AA{} and ${\varepsilon _{\rm OO}} / k =78.23$~K.\label{fluid}}
}%
\hfill%
\parbox[t]{0.5\textwidth}{%
\caption{(Color on-line) Behavior of the axial profiles for the normalized screening strength
axial profiles, $\Im (z)\equiv 1- ( \sigma (z) /  \sigma _{\rm s})$, for the
graphene/fluid interface when the fluid is either pure water at zero surface
charge or an electrolyte-polyelectrolyte solution with a surface-charge range
of $0<\sigma _{\rm s} (\rm C~m^{-2} )\leqslant 0.1525$.\label{strength}}
}%
\end{figure}


The re-structuring of the interfacial region can be interpreted in terms of
the axial profiles of the corresponding electric fields,  $E(z)$, as
illustrated in figure~\ref{fluid} where we should note that $\sigma _{\rm s}
={E(0)\mathord{\left/ {\vphantom {E(0) 4\pi }} \right.
\kern-\nulldelimiterspace} 4\pi } $ is represented by the horizontal lines
($z<2.6$~\AA), and that $0\cong \left[\sigma (z_{\otimes } )-\sigma _{\rm s}
\right]{\rm \lesssim }1.52$ where 2.6~{\AA} ${\rm \lesssim} \,\, z_{\otimes } \,\, {\rm \lesssim }$ 2.92~{\AA} \, when $0\leqslant \sigma _{\rm s} (\rm C~m^{-2} )\leqslant  0.1525$. In this figure we again observe the appearance of a third peak
at $\sim$ 4--4.5~\AA \, from the graphene surface, which moves closer to the
surface as the graphene surface-charge increases in response to the stronger
wall-polyion interactions and the related counterion condensation.  However,
the most revealing feature in this plot is the size of the electric field
``barrier'' $\left[E(z_{\otimes } )-E(0)\right]_{\sigma _{\rm s} } $ represented by
the first peak of $E(z)$ that signifies the occurrence of the SCA phenomenon.
Interestingly, the oscillatory behavior of $E(z)$  translates into
an SCA phenomenon even for the case of an uncharged surface, $\sigma _{\rm s} =0$, for which the electric field ``barrier'' $\left[E(z_{\otimes }
)-E(0)\right]_{\sigma _{\rm s} =0} $ takes the largest value.  Yet, a more
revealing way to look at the SCA and related phenomena is by plotting the
normalized screening strength $\Im (z)$ in figure~\ref{strength}, and observe
the three distinct regions described by equations~\eqref{eq7}--\eqref{eq9}. Note that the SCA signature in our system is revealed by the prominent first valley $\Im
(z_{\otimes } )<0$ that disappears as $\sigma _{\rm s} \to 0.1525~{\rm C~m}^{-2} $.

\begin{wrapfigure}{l}{.53\textwidth}
\centering
\vspace{-4pt}
\includegraphics
[width=0.51\textwidth]{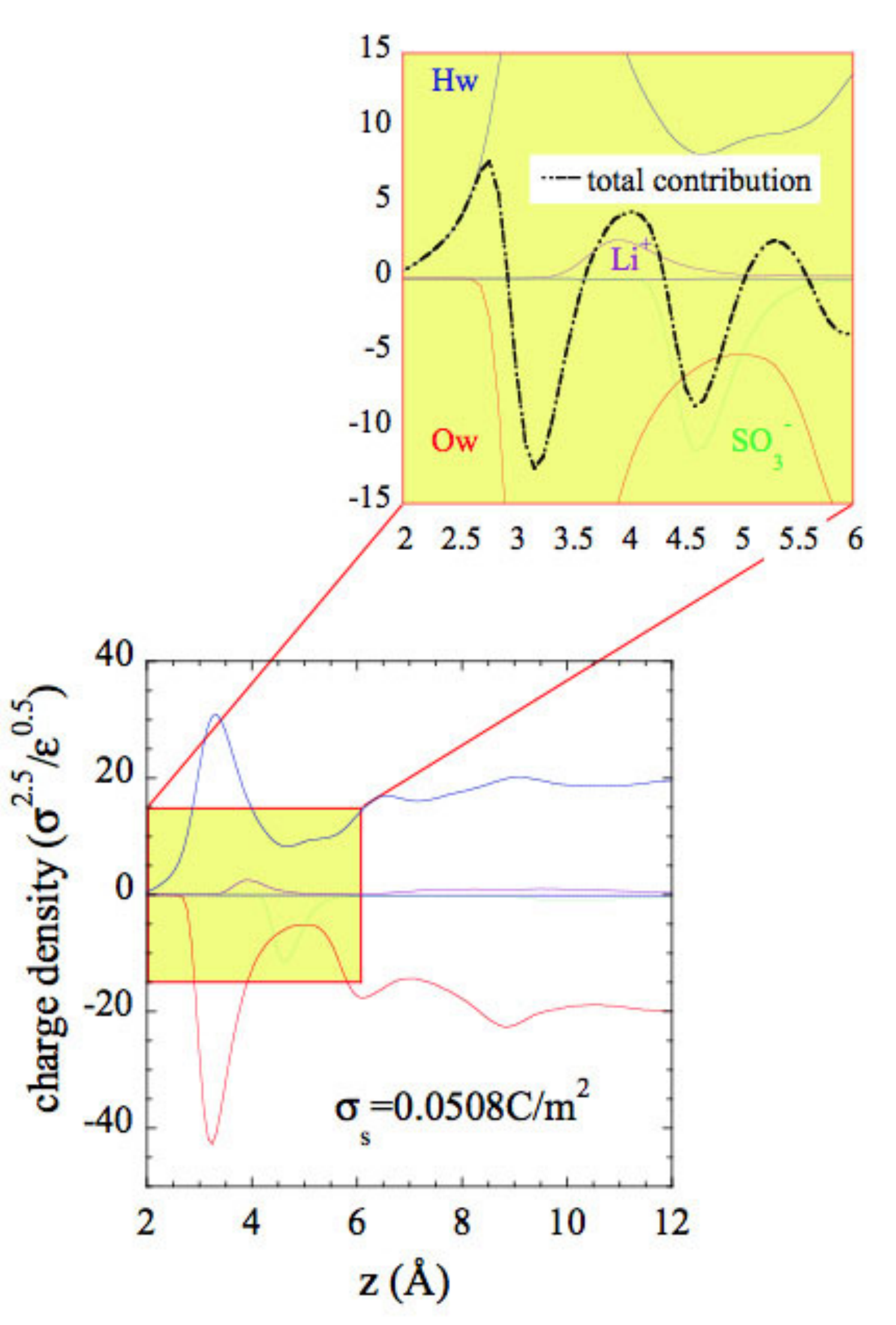}
\vspace{-.15in}
\caption{(Color on-line) Individual species contributions, $q_{i} \rho _{i} (z)$, to the
axial profiles of the charge density, $\rho _{Q} (z)$,     for the
graphene/fluid interface when the fluid is an electrolyte-polyelectrolyte
solution with a surface-charge of $\sigma _{\rm s} =
0.0508~{\rm C~m}^{-2}$.\label{species}}
\vspace{0pt}
\end{wrapfigure}

The results in figures~\ref{charge}--\ref{strength} suggest that the observed
SCA must be associated with the competing adsorption of water, hydrated ions
and poly-ions onto the graphene layer.  This competition is obviously absent
in the case of primitive models where the continuum solvent plays the passive
role of a dielectric screener of the electrostatic interactions,
i.e., there is no inhomogeneous interfacial
solvent distribution due to its interaction with the solid phase.  In order to
shed additional light onto the SCA mechanism, and concomitant CR and CI, in
figure~\ref{species} we plot the individual contribution of each charged site to the charge-density profile for the representative surface-charge $\sigma _{\rm s} =0.0508$~C~m$^{-2} $.
From the zoomed in portion of this picture it becomes
clear that the oxygen and hydrogen sites of the first layer of adsorbed water
are the sole contributors to the first peak and valley of the charge-density
profile, and consequently, to the SCA peak in figure \ref{fluid}.  Note also
that the hydrated lithium ion makes an additional positive contribution to the
second peak of the charge-density profile $\rho _{Q} (z)$, whose main
contribution comes from the balance between the oxygen and hydrogen sites of
the innermost (adsorbed) water layer.  Moreover, while the sulfonate groups
from the innermost PE backbones make the main contribution to the second
valley of $\rho _{Q} (z)$, a balance between the charged sites of water and
the sulfonate groups define almost exclusively the third peak.

\section{Concluding remarks}
\label{sec:Concluding}

We have performed a molecular dynamics simulation study of the interfacial
behavior of short-chain lithium poly\-styrene-sulfonate aqueous solutions in
contact with positive-charged graphene surfaces, in the presence of barium
chloride, to highlight the central role that the discrete nature of the
solvent plays in the occurrence of surface charge amplification.  In contrast
to other recent studies~\cite{GarciaSoft5,GarciaJour6,Messina7,Wang8,Wang9,Yu10,Ravindran18},
here we have used explicit and realistic descriptions of the solvent water,
the chain backbones, and the ionic species in solution that allowed us to
unravel the mechanism underlying the screening of the surface charge by the
surrounding aqueous environment.

\begin{wrapfigure}{l}{.55\textwidth}
\centering
\vspace{-0.6cm}
\includegraphics
[width=0.5\textwidth]{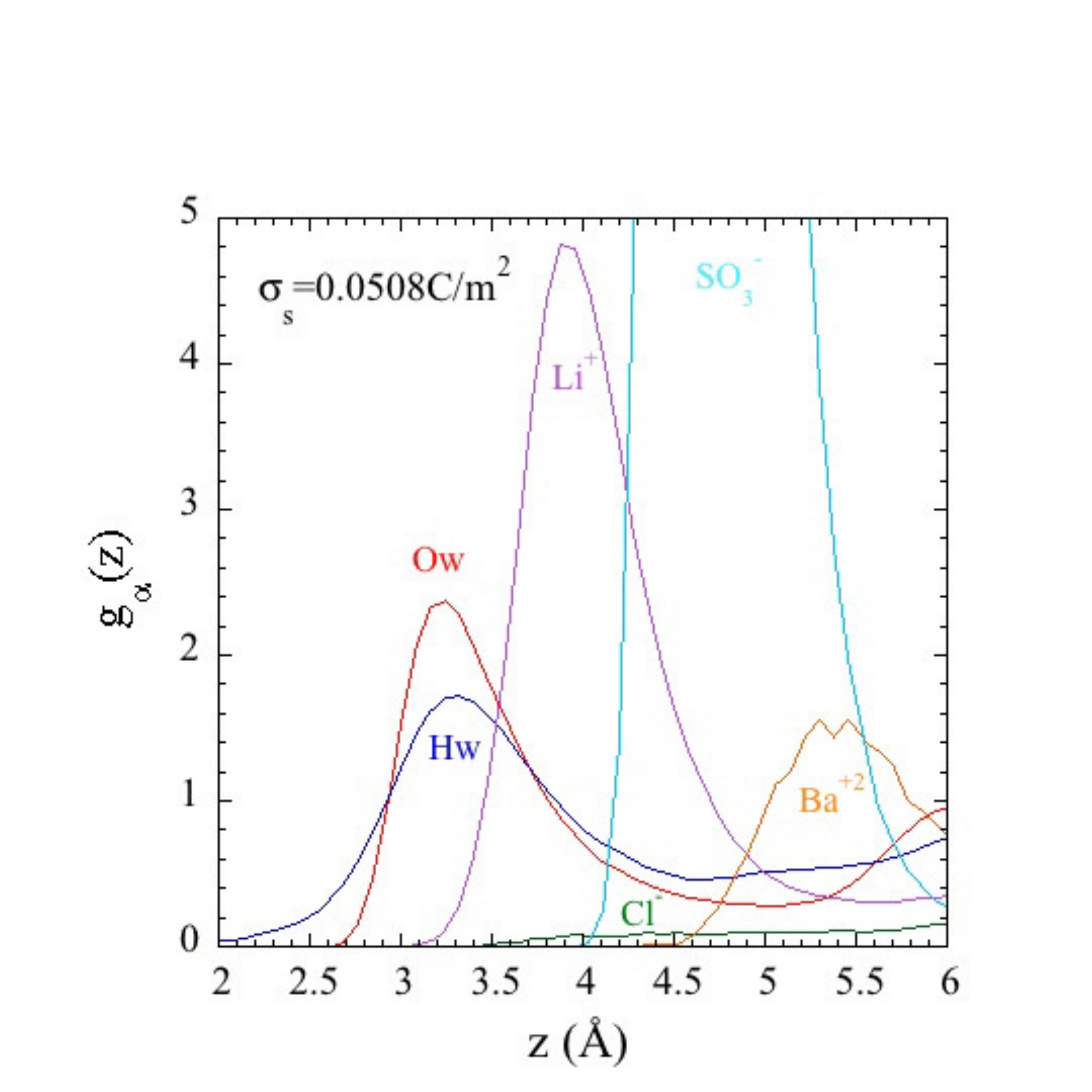}
\caption{(Color on-line) Individual species axial distribution functions, $g_{i} (z)=\rho _{i}
(z) /  \rho _{i}^{\rm bulk}$, for the graphene/fluid interface when the fluid is
an electrolyte-polyelectrolyte solution with a surface-charge of $\sigma _{\rm s}
=0.0508~{\rm C~m}^{-2} $.\label{functions}}
\end{wrapfigure}

In particular, the simulated interfacial structures indicate that the SCA
resulting from the adsorption of short-chain polystyrene-sulfonates onto
charged surfaces originates in the induced orientational structure of the
interfacial water, where the water's hydrogens and oxygens lay in parallel
planes with opposite density charges, aided by the adsorbed polyelectrolyte
chains and condensed counterions, though contributing only marginally by
enforcing the system electroneutrality. Consequently, the magnitude of the
SCA depends mainly on the relative overlapping of the distribution of the
hydrogen- and oxygen-site charges (see figure~\ref{functions}), which in turn,
is controlled by the tilt angle $\theta (z)$ (figure~\ref{axial}).  In other
words, the solvent (water), rather than the ionic species, becomes the major
player in the SCA phenomenon; water mediates the interactions of all species
with the charged surface, i.e., the ions are adsorbed as hydrated
species.  This scenario highly contrasts with that from SCA phenomena observed
by theory~\cite{Jimenez1,GarciaSoft5,GarciaJour6}
and simulation~\cite{Wang8,Wang9,Ravindran18,Tanaka19}
involving primitive models, where a continuum dielectric describes the solvent
in the screening of the Coulombic interactions.  Yet, we can draw some common
features between the two scenarios by recognizing that when the solvent water
is described atomistically as an electroneutral set of bonded charged-sites,
it plays the same role as any charged species in the system, except for its
ability to solvate the species due to its polar nature that hinders other
species from approaching the surface. Consequently, while our analysis
suggests an SCA mechanism similar to that proposed originally by
Jimenez-Angeles and Lozada-Cassou~\cite{Jimenez1} ({e.g.,} see insets of
their figure~\ref{axial}), we must highlight some significant conceptual
differences with the latter, namely: (a) the magnitude and asymmetry of the
charges involved in our system are much smaller, i.e., $q_{\rm H_{w} }
=0.4238e=-0.5q_{\rm O_{w} } $ in comparison to $q_{\rm M} =-40e=-20q_{+} =40q_{-} $ in
reference~\cite{Jimenez1}, resulting in a charge asymmetry of 2 compared to 20
in reference~\cite{Jimenez1}, and (b) in our case the partial charges on the
sites of the electroneutral polar solvent are linked by the intramolecular
$\rm O-H$ bonds, i.e., they are correlated by construction.

While we have tackled the specific case of SCA when $\sigma _{\rm s} >0$, a
similar phenomenon can in principle happen for $\sigma _{\rm s} <0$
(e.g., see figure~4 (c)--(d) of Wang et al.~\cite{Wang8}, and
figure~4~(b) of Tanaka and Grosberg~\cite{Tanaka19} which was not recognized as
such until recently~\cite{Messina7}).  Following a similar analysis as for
equations~\eqref{eq7}--\eqref{eq9}, their counterparts for the case of $\sigma _{\rm s} <0$
become,

(a) surface-charge amplification (SCA), i.e.,
\begin{equation} \label{eq10}
\left|\sigma _{\otimes } \right|>\left|\sigma _{\rm s} \right|>0{\rm \; \; \; }\to {\rm \; \; }\int _{0}^{z_{\otimes } }\rho _{Q} (z)\rd z <0{\rm \; \; }\to {\rm \; }\Im (z_{\otimes } )<0;
\end{equation}

 (b) charge inversion (CI), i.e.,
\begin{equation} \label{eq11}
\left|\sigma _{\rm s} \right|>\left|\sigma _{\otimes } \right|>0{\rm \; \; \; }\to {\rm \; \; }\left|\sigma _{\rm s} \right|>\int _{0}^{z_{\otimes } }\rho _{Q} (z) \rd z {\rm \; \; }\to {\rm \; }\Im (z_{\otimes } )<1;
\end{equation}

(c) charge reversal (CR), i.e.,
\begin{equation} \label{eq12}
\sigma _{\otimes } >0{\rm \; \; \; }\to {\rm \; \; }\left|\sigma _{\rm s} \right|<\int _{0}^{z_{\otimes } }\rho _{Q} (z)\rd z {\rm \; \; }\to {\rm \; }\Im (z_{\otimes } )>1.
\end{equation}

Thus, according to equations~\eqref{eq7}--\eqref{eq9} and ~\eqref{eq10}--\eqref{eq12}, the general conditions for the occurrence of SCA, CI and CR for an arbitrary surface charge $\sigma _{\rm s} \,{\gtrless }\, 0$ can be described as follows,

\begin{eqnarray} \label{eq13}
\left(\sigma _{\otimes }  \sigma _{\rm s} \right)>0& {\text{with} }\left|\sigma _{\otimes } \right| >\left|\sigma _{\rm s} \right| &\to {\rm SCA},\\
\label{eq14}
\left(\sigma _{\otimes }  \sigma _{\rm s} \right)>0& {\text{with}}\left|\sigma _{\otimes } \right| <\left|\sigma _{\rm s} \right| &\to {\rm CI},\\
\label{eq15}
\left(\sigma _{\otimes }  \sigma _{\rm s} \right)<0& {\text{with}}\left|\sigma _{\otimes } \right| \;{\rm \gtrless}\left|\sigma _{\rm s} \right| &\to {\rm CR}.
\end{eqnarray}

We should mention that Tanaka and Grosberg~\cite{Tanaka19} analyzed the ion
adsorption/condensation onto macroion surfaces by molecular dynamics of
primitive models and described the over-screening of the negative-charged
macroion as a ``giant charge inversion''.  In particular, we note that their
figure~\ref{axial} dealing with a ${Q(r\approx 3a)\mathord{\left/ {\vphantom
{Q(r\approx 3a) \left|Q_{0} \right|}} \right. \kern-\nulldelimiterspace}
\left|Q_{0} \right|} \approx 2.5$ and $Q_{0} =-28e$ does not portray a CI as
claimed but a CR phenomenon, i.e., $-{Q(r\approx 3a)\mathord{\left/
{\vphantom {Q(r\approx 3a) Q_{0} }} \right. \kern-\nulldelimiterspace} Q_{0} }
\sim \Im (3a)>1$ according to equation~\eqref{eq12}, where the same conclusion can be
reached by Guerrero-Garcia et al.'s~\cite{GarciaSoft5} analysis
involving their equation~(5).  Likewise, the behavior described in Tanaka and
Grosberg's  figure~4~(b) for ${Q(r\approx 8a)\mathord{\left/ {\vphantom
{Q(r\approx 8a) \left|Q_{0} \right|}} \right. \kern-\nulldelimiterspace}
\left|Q_{0} \right|} \approx -\, 0.6$ and $Q_{0} =-28e$ does not describe a CI
but an SCA phenomenon because ${-\, Q(r\approx 8a)\mathord{\left/ {\vphantom
{-\, Q(r\approx 8a) Q_{0} }} \right. \kern-\nulldelimiterspace} Q_{0} } \sim
\Im (8a)<0$ according to equation~\eqref{eq12} and in agreement with Guerrero-Garcia
et al.'s analysis~\cite{GarciaSoft5}.

Our analysis should be put in the context of, and in contrast against, the
recent developments dealing mainly with primitive models.  Since we are
dealing with an explicit atomistic description of the solvent, our results
point immediately to some yet unexplored issues including, (a) the potential
effect of the water's flexible geometry on the strength of its adsorption on
the grapheme surface and on the resulting charge screening, (b) the effect of
confinement as a result of the overlapping of two approaching aqueous-graphene
interfaces, and (c) the existence of a similar mechanism for the negative SCA
involving the corresponding poly-cation species.  Note that the proposed SCA
mechanism is in principle independent of the involved water model, though the
strength (and width of the surface-charge window in which SCA occurs) might
depend on the water model chosen.  In particular, we should expect a clear
model dependence when comparing the simulation results from a rigid against its
flexible geometry counterpart.

\section*{Acknowledgements}

Research sponsored by the Division of Chemical Sciences, Geosciences, and Biosciences, Office of Basic Energy Sciences, U.S. Department of Energy.

\newpage

\ukrainianpart

\title{Молекулярний механізм збільшення поверхневого заряду і споріднені явища на межі розділу водний полі\-електро\-літ-графен}

\author{А.А. Кіалво\refaddr{label1}, Дж.М. Сімонсон\refaddr{label2}}

\addresses{
\addr{label1}Відділ хімічних наук,
Національна лабораторія Оук Рідж, Оук Рідж, Тенессі, США
\addr{label2}Відділ досліджень нейтронним розсіянням,
Національна лабораторія Оук Рідж, \\ Оук Рідж, Тенессі, США}

\makeukrtitle

\begin{abstract}
\tolerance=3000%
У статті нами розглянуто недавно виявлене явище збільшення поверхневого заряду (ЗПЗ) [Jimenez-Angeles~F. and Lozada-Cassou~M., J.~Phys. Chem.~B, 2004, {\bf 108}, 7286]. Методом молекулярної динаміки проведено моделювання водних розчинів електролітів з багатовалентними
катіонами, що перебувають у контакті з графеновими стінками;
воду в моделюванні представлено явно з використанням реалістичної молекулярної моделі. Нами показано,
що явище ЗПЗ в таких системах, на відміну від примітивних моделей, не включає ні контактної
коадсорбції негативно заряджених макроіонів, ні двовалентних катіонів з великою асиметрією у розмірах та зарядах, як це вимагається у випадку моделей з розчинниками, представленими
як суцільне середовище. Насправді ефект ЗПЗ проявляється за рахунок вибіркової адсорбції води
(через гідратовані іони) з напрямком диполів, орієнтованих таким чином, щоб підсилювати, а не
екранувати позитивно заряджену поверхню графену.

\keywords молекулярне моделювання, межа розділу рідина-тверде тіло,
водні полі\-електроліти, збільшення поверхневого заряду, зарядова інверсія і реверсування

\end{abstract}


\newpage

\begin{thebibliography}{19}

\bibitem{Jimenez1} Jimenez-Angeles~F. and Lozada-Cassou~M.,
J.~Phys. Chem.~B, 2004, {\bf 108}, No.~22, 7286;\\
\doi{10.1021/jp036464b}.

\bibitem{Sjostrom2} Sjostrom~L., Akesson~T., and Jonsson~B.,
Ber. Bunsen Ges. Phys. Chem., 1996, {\bf100}, No.~6, 889;\\
\doi{10.1002/bbpc.19961000634}.

\bibitem{hansen3} Hansen J.P. and Lowen~H., Annu. Rev. Phys. Chem., 2000, {\bf51},
209;\\
\doi{10.1146/annurev.physchem.51.1.209}.

\bibitem{Decher4} Decher~G. and Hong~J.D.,
Makromol. Chem. Macromol. Symp., 1991, {\bf46}, 321;\\
\doi{10.1002/masy.19910460145};
%
Dan~N., Nano Lett., 2003, {\bf3}, No.~6, 823;
\doi{10.1021/nl034122b};\\
%
Pittler~J., Bu~W., Vaknin~D., Travesset~A., McGillivray~D.J., and
Loesche~M., Phys. Rev. Lett., 2006, {\bf97}, No.~4, 046102;
\doi{10.1103/PhysRevLett.97.046102}.

\bibitem{GarciaSoft5} Guerrero-Garcia~G.I., Gonzalez-Tovar~E., and de la Cruz~M.O.,
Soft Matter, 2010, {\bf6}, No.~9, 2056;\\ \doi{10.1039/B924438G}.

\bibitem{GarciaJour6} Guerrero-Garcia~G.I., Gonzalez-Tovar~E., Chavez-Paez~M.,
and Lozada-Cassou~M., J.~Chem. Phys., 2010, {\bf 132}, No.~5,
054903; \doi{10.1063/1.3294555}.

\bibitem{Messina7} Messina~R., J.~Chem. Phys., 2007, {\bf127}, No.~21, 214901;
\doi{10.1063/1.2807228}.

\bibitem{Wang8} Wang~Z.-Y. and Ma~Y.-Q., J.~Phys. Chem.~B, 2010, {\bf114}, No.~42, 13386;
\doi{10.1021/jp106118q}.

\bibitem{Wang9} Wang~Z.-Y. and Ma~Y.-Q., J.~Chem. Phys., 2010, {\bf133}, No.~6,
064704; \doi{10.1063/1.3469795}.

\bibitem{Yu10} Yu~J., Aguilar-Pineda~G.E., Antillon~A., Dong~S.H., and Lozada-Cassou~M.,
J.~Colloid Interface Sci., 2006, {\bf295}, No.~1, 124;
\doi{10.1016/j.jcis.2005.08.016}.

\bibitem{Chialvo11} Chialvo~A.A. and Simonson~J.M.,
J.~Phys. Chem.~C., 2008, {\bf 112}, No.~49, 19521;
\doi{10.1021/jp8041846}.

\bibitem{Gossl12} Gossl~I., Shu~L.J., Schluter~A.D., and Rabe~J.P.,
J.~Am. Chem. Soc., 2002, {\bf 124}, No.~24, 6860;\\
\doi{10.1021/ja017828l};
%
Maiti~P.K. and Bagchi~B., Nano Lett., 2006, {\bf6}, No.~11, 2478;\\
\doi{10.1021/nl061609m};
%
Lyulin~S., Darinskii~A., and Lyulin~A., e-Polym., 2007, No.~097-1/14;
%
Gillies~G., Lin~W., and Borkovec~M., J.~Phys. Chem.~B, 2007,
{\bf111},  No.~29, 8626; \doi{10.1021/jp069009z};
%
\\ Kundagrami~A. and Muthukumar~M., J.~Chem. Phys., 2008, {\bf128},
No.~24, 244901;\\ \doi{10.1063/1.2940199};
%
May~S., Iglic~A., Rescic~J., Maset~S., and Bohinc~K., J.~Phys.
Chem.~B, 2008, {\bf112}, No.~6, 1685; \doi{10.1021/jp073355e}.

\bibitem{Grosberg13} Grosberg~A.Y., Nguyen~T.T., and Shklovskii~B.I.,
Rev. Mod. Phys., 2002, {\bf74}, No.~2, 329;\\
\doi{10.1103/RevModPhys.74.329}.

\bibitem{Joanny14} Joanny~J.F., Eur. Phys.~J.~B, 1999, {\bf9}, No.~1, 117;
\doi{10.1007/s100510050747}.

\bibitem{Dobrynin15} Dobrynin~A.V., Deshkovski~A., and Rubinstein~M.,
Macromolecules, 2001, {\bf34}, No.~10, 3421;
\\ \doi{10.1021/ma0013713}.

\bibitem{Diehl16} Diehl~A. and Levin~Y., J.~Chem. Phys., 2006, {\bf125},  No.~5, 054902;
\doi{10.1063/1.2222372};\\
%
Pianegonda~S., Barbosa~M.C., and Levin~Y., Europhys. Lett., 2005,
{\bf71}, No.~5, 831; \\ \doi{10.1209/epl/i2005-10150-y}.

\bibitem{vonSeggern17} von Seggern~D., Standard Curves and Surfaces.
CRC Press, Boca Raton, 1993.

\bibitem{Ravindran18} Ravindran~S. and Wu~J.Z., Langmuir, 2004, {\bf20}, No.~17, 7333;
\doi{10.1021/la0493619};\\
%
Messina~R., Holm~C., and Kremer~K., Langmuir, 2003, {\bf19},
No.~10, 4473; \doi{10.1021/la026988n}.

\bibitem{Tanaka19} Tanaka~M. and Grosberg~A.Y., J.~Chem. Phys., 2001, {\bf115}, No.~1,
567; \doi{10.1063/1.1377033}.

\end{thebibliography}
\end{document}